\documentclass[pre,aps,twocolumn,superscriptaddress,longbibliography,floatfix,notitlepage]{revtex4-2}

\usepackage[utf8x]{inputenc}
\usepackage[english]{babel}
\usepackage[T1]{fontenc}
\usepackage{lmodern}

\usepackage{amsmath,amsfonts,amssymb,amsthm,bm,times,dcolumn}

\usepackage{microtype}
\usepackage{braket}  
\usepackage{gensymb} 
\usepackage{physics} 

\usepackage[colorlinks={true}, citecolor={blue}, filecolor={blue}, linkcolor={blue}, urlcolor={blue}]{hyperref}
\usepackage{graphicx,color}
\usepackage[caption=false]{subfig}

\begin{document}
\preprint{APS/123-QED}
\title{Effects of Graph Network Connections on The Efficiency of Quantum Annealing}
\author{Atalay Ege}
\email{egea16@itu.edu.tr}
\affiliation{Department of Physics Engineering, \.{I}stanbul Technical University, Sariyer, \.{I}stanbul, 34469, Turkey}
\author{\"{O}zg\"{u}r E. M\"{u}stecapl{\i}o\u{g}lu}
\email{omustecap@ku.edu.tr}
\affiliation{Department of Physics, Ko\c{c} University, Sar{\i}yer, \.Istanbul, 34450, Turkey}
\affiliation{T\"{U}B\.{I}TAK Research Institute For Fundamental Sciences, 41470 Gebze, Turkey}
\date{\today}
\begin{abstract}
Graph structure of quantum spin networks plays an essential role in applying quantum annealing (QA). The Ising model is the typical choice to describe the interactions between the spins in the networks. Here, we explore the interplay of different interaction types between the spins and the structure of the underlying network on the performance of QA. Specifically, we consider the Heisenberg XY and antisymmetric anisotropic exchange interaction models, in addition to the usual Ising Model, in graph structures ranging from linear chains to complete graphs with different connectivities. We find that the QA performance of the Ising model shows less variation with the change of graph structure among different interaction types. The Heisenberg XY model suffers from significant performance losses with increasing connections beyond linear chains, which we attribute to increasing entropy built-up. Anisotropic exchange interactions turn out to be an impractical choice for complete graphs due to the diminishing energy gap during annealing. Furthermore, we investigate the effect of inhomogeneous couplings and reveal a trade-off between coupling strength and performance, especially significant for the case of the anisotropic exchange model.
\end{abstract}
\maketitle
\section{\label{sec:Intro}Introduction}
Gaining insight into complex physical systems is a difficult task and has created the idea of simulating physics with computers~\cite{feyn82}. While classical computers are completely capable, their limitations pushed the research interest in quantum computers.  We will be focusing on the architecture of quantum annealers. They are usually designed to solve optimization problems by mapping them to (mostly Ising) interacting spin network models~\cite{Kado1998,Santoro2002}. A practical challenge of implementing spin networks on quantum annealing hardware is the connectivity of physical qubits~\cite{coupling2020}. Accordingly, the classification of spin network topologies and spin-spin interactions for efficient quantum annealing is a question of both fundamental and practical significance.

Simulating a physical system in the context of quantum computation means that an initial state gets transformed step by step by unitary operators~\cite{deutsch_quantum_1989}. Quantum annealing (QA) differs as the procedure slowly varies a Hamiltonian where the main concern is its ground state.  This approach makes the computation part of the job easier to formulate, and it was also shown that quantum annealing is as strong as the quantum circuit based quantum computation~\cite{Aharonov2008}. Reviews on the subject are available~\cite{Arnab2008,Albash2018} and come with alternative names such as adiabatic quantum computation (AQC) and analog quantum computation. Applications include data analysis~\cite{Hashizume2015}, optimization problems mapped into spin-glasses~\cite{Stoll2019,fullyconnected}, and graph network problems~\cite{graphiso,disorgraph}. D-Wave annealers and their chimera architecture Ising model spin networks offer the state of the art solutions~\cite{johnson_quantum_2011,dwave21}.

In this work, we investigate how different quantum graph networks affect quantum annealing while mapped into different models such as the Ising model, Heisenberg XY model, and antisymmetric anisotropic model spin-spin interactions. In addition, we examine, how those models and different networks respond to inhomogeneities or additional Hamiltonian terms. We work with exact Hamiltonians and their eigenvalues/eigenvectors found with diagonalization. Using energy gap, entanglement entropy, mean magnetization, coherence, and fidelity measures, we quantify the performance of the adiabatic procedure. We find that any network selection for the Ising model is a good candidate for efficient annealing while the same can't be said about Heisenberg and anisotropic models. For the later models, the issue was with the relatively higher entropy build-up for the square type graphs that have higher connectivity than chains. We also check the inhomogeneous coupling cases for all models and while in general, they decrease the efficiency of annealing, we also observe that weaker couplings can help with inefficient network structures. Therefore, our results highlight the importance of network selection for different models regarding optimal solutions.

The structure of the paper is as follows. We introduce the computational process and underlying principles of AQC in Sec.~\ref{sec:model}. Sec.~\ref{sec:results} presents our findings regarding the graph structure dependence of the AQC with Ising model (Sec.~\ref{sec:subsection1}), Heisenberg XY model (Sec.~\ref{sec:subsection2}), and antisymmetric anisotropic model (Sec.~\ref{sec:subsection3}). While those sections cover the different coupling schemes between qubits, we examine the case of inhomogeneous couplings in Sec.~\ref{sec:subsection4}. Finally, we conclude in Sec.~\ref{sec:conc}.

\section{\label{sec:model} Model}
AQC is based upon the quantum adiabatic theorem~\cite{Born28},  according to which a system prepared in the ground state of a relatively simple Hamiltonian remains in the ground state as the Hamiltonian is varied adiabatically slow towards a more complex target Hamiltonian. In mathematical terms, we write
\begin{equation}
    H(\lambda) = \lambda H_1+(1-\lambda)H_0, 
    \label{eq:adia1}
\end{equation}
where, $\lambda$ is the adiabatic parameter in the domain $[0,1]$ controlling the speed of procedure. $H_0$ is the simpler Hamiltonian where the system starts off in its ground state and then reaches the target Hamiltonian $H_1$ with $\lambda$ dependence. Quantum annealing systems typically consist of N  identical two-level systems (qubits), which can be described in terms of spin-$1/2$ particles. We scale the parameters of the Hamiltonians
with the energy of the spins and express the simple Hamiltonian $H_0$ as
\begin{equation}
    H_0 = -\sum_{i=1}^NS_i^{x}.
    \label{eq:Hamground}
\end{equation}
Here, $S_i^{\sigma}$, $\sigma=x,y,z$ are components of Pauli spin-$1/2$ operators for the spin at site $i$. We consider three target Hamiltonians that can be found in the literature, first of which is the Ising model~\cite{Nishi2001}
\begin{equation}
    H_1^{\text{IM}} = \sum_{i\neq j}^NJ_{ij}S_i^zS_j^z+\sum_{i=1}^Nh_iS_i^z,
    \label{eq:hamIs}
\end{equation}
second one is the Heisenberg XY model~\cite{Mattis1985}

\begin{equation}
    H_1^{\text{HM}} = \sum_{i\neq j}^NJ_{ij}(S_i^xS_j^x+S_i^yS_j^y)+\sum_{i=1}^Nh_iS_i^z,
    \label{eq:hamHei}
\end{equation}
and the final one is the antisymmetric anisotropic exchange~(Dzyaloshinskii-Moriya) interaction~\cite{DM,magnetismcondensed}

\begin{equation}
    H_1^{\text{DMM}} = \sum_{i\neq j}^NJ_{ij}(S_i^zS_j^x-S_i^xS_j^z)+\sum_{i=1}^Nh_iS_i^z.
    \label{eq:hamAn}
\end{equation}
Here, $h_i$ is the magnetic field at the spin site $i$; $J_{ij}$ are the spin-spin coupling coefficients. Couplings will be ferromagnetic if they are negative and anti-ferromagnetic if they are positive.

Performance of QA can be quantified in terms of figure of merits that can be simply determined using the eigenvalues of the Hamiltonian~(\ref{eq:adia1}) for a given $\lambda$. The first of which is the energy gap
\begin{equation}
    \Delta E  = E_1-E_0,
    \label{eq:engap}
\end{equation}
determines the validity and performance of adiabatic processes~\cite{Santoro2002} according to quantum adiabatic theorem~\cite{Born28}. 

The other figure of merit is the Von Neumann entanglement-entropy of the system~\cite{entanglement}
\begin{equation}
    S = -\sum_i\sigma_i^2\log(\sigma_i^2),
    \label{eq:entropy}
\end{equation}
where, the $\sigma_i$ are the non-zero singular values of the ground state. They can be found with partitioning the system 
\begin{equation}
    \Psi = \sum_{i,j}A_{ij}\ket{i}\otimes\ket{j},
    \label{eq:partition}
\end{equation}
where, $\ket{i}$ and $\ket{j}$ bases depend on how do we partition the system. The resulting $A$ matrix is used to find singular values via singular value decomposition~\cite{svd}. The number of singular values has implications over the entanglement level of the system.

Another figure of merit is the average magnetization
\begin{equation}
    m = \frac{1}{N}\sum_i\langle S_i^z\rangle,
\end{equation}
where, $\langle S^z_i\rangle=\bra{\psi_0(\lambda)}S^z_i\ket{\psi_0(\lambda)}$. Here, $\psi_0(\lambda)$ is the ground state of the Eq.~(\ref{eq:adia1}).

Coherence in the ground state can affect the performance of the quantum annealing and characterized here using the~\cite{Coherence}
\begin{equation}
    C_{l_1}=\sum_{i\neq j}|\rho_{ij}|,
    \label{eq:coh}
\end{equation}
where, $\rho_{ij}$ are the off-diagonal elements of the ground state density matrix
\begin{equation}
    \rho = \ket{\psi_0(\lambda)}\bra{\psi_0(\lambda)}.
    \label{eq:rho}
\end{equation}

The final figure of merit is the fidelity. It measures how close we are to the ground state of the target Hamiltonian during the computation. To compute fidelity we will follow Jozsa's axioms~\cite{Jozsa,FidRev} and use the form
\begin{equation}
    \mathcal{F}(\nu,\rho)=\textup{tr}(\nu\rho).
    \label{eq:Fide}
\end{equation}
Here, $\nu$ is the ground state density matrix of the target Hamiltonian. We can use this definition of fidelity as the ground state of the target Hamiltonian will be a pure state for any scheme we are working with.

\begin{figure}[h!]
\includegraphics[width=\linewidth]{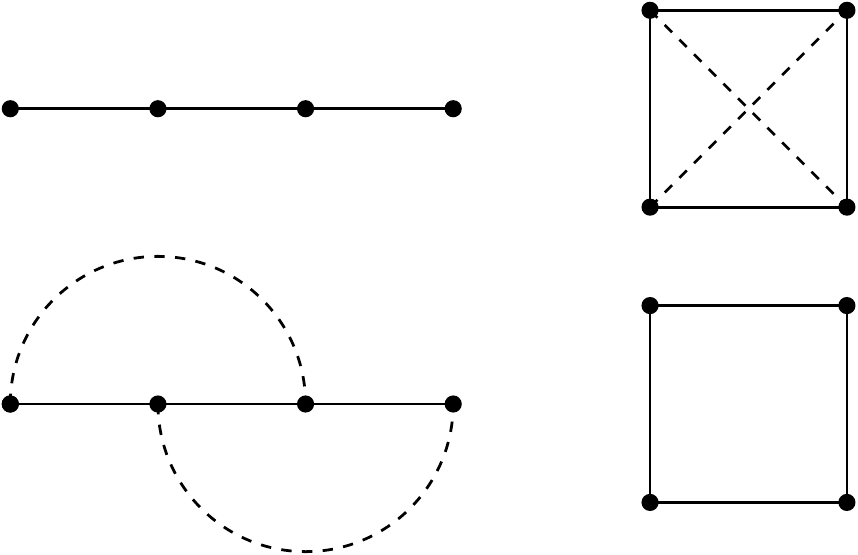}
\caption{\label{fig:graphs}Graph structures of quantum networks that we consider in this work. Vertices are the spin qubits and solid/dashed lines represent the interaction between qubits. They are named as: linear chain (top left), linear chain with loops (bottom left), square graph (bottom right), complete square graph (top right). }
\end{figure}

Those definitions are sufficient to check the performance of the AQC. To build Hamiltonians corresponding to different models, we consider four different networks with four qubits presented in Fig.~\ref{fig:graphs}. All four graphs have four nodes. In terms of their number of edges they are ordered as linear graph (3 edges), square graph (4 edges),  linear graph with loops (5 edges), complete square graph (6 edges). Linear chain with loops can also be envisioned as an incomplete square graph.


\section{\label{sec:results}Results}

\subsection{\label{sec:subsection1} Ising Model}

We start to compare behaviors of different quantum network structures in the Ising model. The first figure of merit we check is the energy gap. The results show that for any graph network structure that shown in Fig.~\ref{fig:graphs}, the energy gap behavior is similar. We observe minimum points in each case that doesn't approach zero. So, there is nothing wrong with the adiabaticity of Ising model quantum networks. 
\begin{figure*}[t!]
    \centering
    \includegraphics[width=\linewidth]{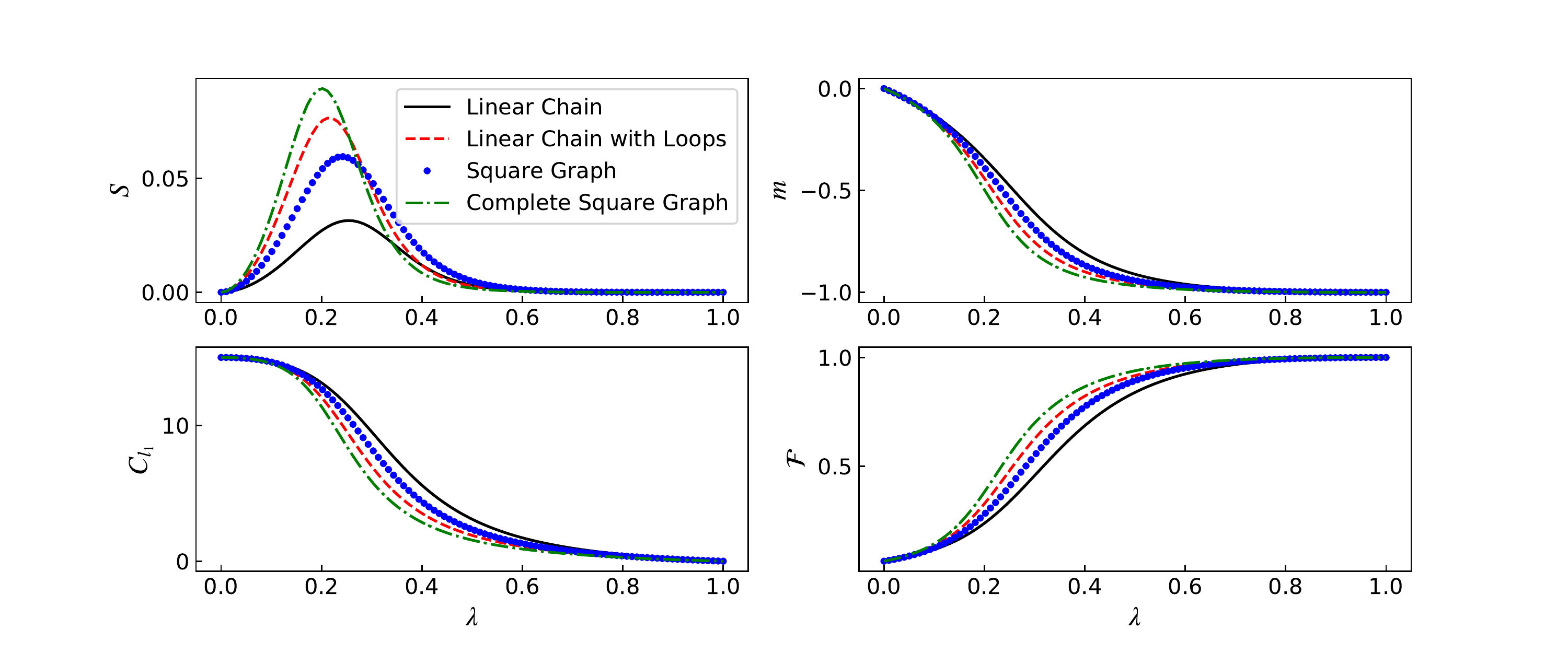}
    \caption{(Color online) The figure of merits that have been calculated for Ising Model spin networks. All quantities are dimensionless and controlled by the adiabatic parameter~($\lambda$). Curves represent the four graph types~(Fig.~\ref{fig:graphs}) that we are comparing, black solid lines for the linear chain, red dashed lines for the linear chain with loops, blue dots for the square graph, and green dash dotted lines for the complete square graph. At the top left panel we can see the evolution of entropy~($S$), top right panel represents the evolution of mean magnetization~($m$), bottom left panel shows the $l_1$ norm coherence~($C_{l_1}$) of the system during the adiabatic procedure and finally bottom right panel gives us the fidelity~($\mathcal{F}$) evolution of the system. $J_{ij}$ couplings were $-1$ and $h_i$ field strengths were $1$, resulting in a $\ket{1111}$ ground state for the target Hamiltonian. Since our calculations are dimensionless, $\hslash/2$ constants were omitted from spin operators.}
    \label{fig:ising}
\end{figure*}

The next figure of merit is the entanglement entropy. We present the results at the top left panel in Fig.~\ref{fig:ising}. The first observation is that the entanglement behavior of all structures is similar but their maximums differ. We see that square type graphs on average have higher levels of entanglement and entropy. Entropy is higher in the spin networks with more edges. Remarkably, the same order of the curves with the number of graph edges for the 
other figure of merits can be seen in other panels in Fig.~\ref{fig:ising}. To be a driving force for quantum annealing even a small amount of entropy is enough~\cite{bauer2015entanglement}. We note that we partition the system to calculate entropy in Eq.~(\ref{eq:partition}). Fig.~\ref{fig:ising}'s partitions are made from the middle of 4 qubits, other ways of doing it involve slicing the structures of Fig.~\ref{fig:graphs} as 1 and 3 qubits or 3 and 1 qubits but the difference is not significant and the general behavior does not change. 

The third figure of merit is the mean magnetization, results of which can be seen at the top right panel of Fig~\ref{fig:ising}. We see that all four quantum networks follow a similar curve towards the mean magnetization of $-1$. The usefulness of those curves depends on what is measurable for the system. It is not always easy to find the probability of target Hamiltonian's ground state in practice. Then, measurements of magnetization can help to declare if the annealing procedure was successful.  In general, we will not know the ground state of the target Hamiltonian, unlike our case where the approach to $-1$ is what we expect. However, even with unknown target states the stability indicated with magnetization curves can be helpful. If we want to optimize for the best quantum network, then the curves that approach the target state faster can be indicative. We find that the complete square graph approaches the $-1$ mean magnetization faster than others. Hence, it should be the top choice for optimizing the problem solely based on magnetization curves.

Fourth figure of merit is the $l_1$ norm coherence that we define with Eq.~(\ref{eq:coh}). The results for it is shown at the bottom left panel of Fig.~\ref{fig:ising}. Coherence behavior is quite similar to magnetization behavior. We see that decoherence happens naturally towards the more stable states of the structures. Initial high coherence is due to the ground state of Eq.~(\ref{eq:Hamground}) where the system starts in. This way, we are guaranteed to have a superposition of states at the beginning of the computation. In our context, the expectation is to not have a very high decoherence rate depending on the adiabatic parameter ($\lambda$) as that would mean we might end up with a false minimum. Also, for quantum computers, coherence time is crucial especially considering practical applications~\cite{CohTime}. Then, we conclude that all quantum networks have sufficient coherence from our results.

The final figure of merit is fidelity. We see the results at the bottom right panel of Fig.~\ref{fig:ising}. High fidelity operations for quantum annealing are important to have in practical applications~\cite{Highfid}. Looking at the results complete square graph shows the highest fidelity on average during the computational process. The fidelity results are directly related to the target state, those can justify the other figure of merits as well. So, our conclusion regarding the mean magnetization is supported by fidelity. We also observe that the higher decoherence rate of the complete square graph is not an issue for performance as long as its fidelity is high. Another remark to make is that the maximum entropy state of the computation doesn't indicate a high fidelity ground state but a high slope point for the magnetization, decoherence, and fidelity curves.

 Ising model is a candidate for successful adiabatic quantum computation for any type of graph structure. The top-performing graph structure is the complete square graph. Therefore, we state that Ising model spin interactions on ladders are working as intended in terms of quantum annealing. However, we can't conclude immediately that the Ising model will be this reliable for lattices or spin glasses that have long-range interactions.

\subsection{\label{sec:subsection2} Heisenberg Coupling}
\begin{figure}[h!]
\includegraphics[width=\linewidth]{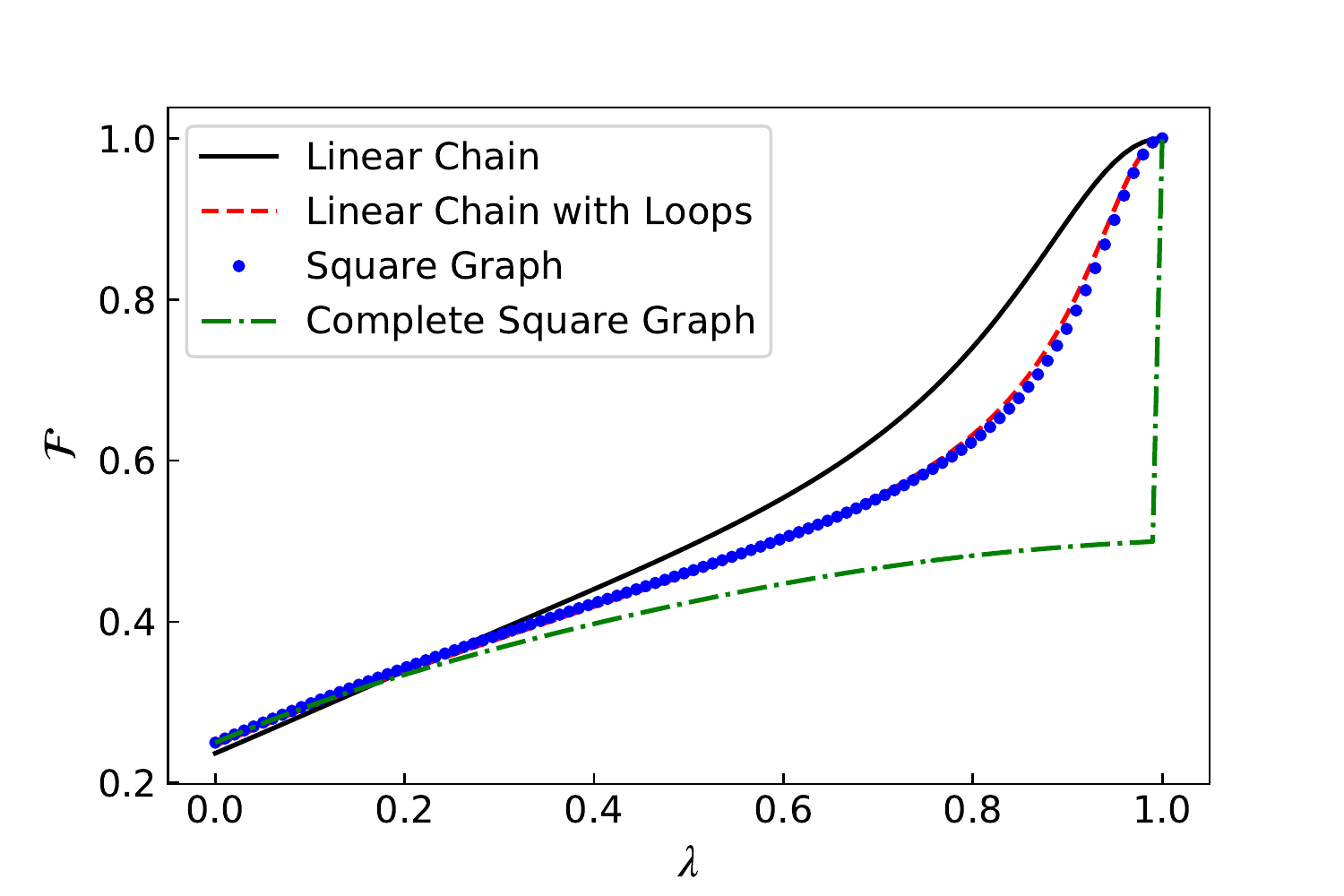}
\caption{\label{fig:hei}(Color Online) Quantum annealing fidelity ($\mathcal{F}$) results for Heisenberg model on different spin network geometries. $\lambda$ is the adiabatic control parameter and black solid lines represent the linear chain, red dashed lines represent the linear chain with loops, blue dots represent the square graph, dash dotted green lines represent the complete square graph. Same as Ising model, $J_{ij}$ coefficients are $-1$ ferromagnetic, $h_i$ field strengths were $1$. }
\end{figure}

We also look for the performance results in the Heisenberg XY model (HM). We follow the same procedure of checking the figure of merits like the Ising model computations. Since we find that mean magnetization and fidelity serve the same purpose there is no need to repeat both of them. Our results for the fidelity in Heisenberg XY model quantum networks are in Fig.~\ref{fig:hei}.

Fidelity results are low during the annealing procedure for any graph structure, especially for the complete square graph. Then, the first observations suggest that Heisenberg model quantum networks are inefficient for AQC.  Without giving the figure here, we can talk about the energy gap and entropy behavior of graph networks as well. It appears none of those structures achieve an early minimum for energy gap and their entropy while high on average doesn't have a maximum point like Ising model results of Fig.~\ref{fig:ising}. Those computations are enough to discredit the structures as efficiently working adiabatic quantum computers considering the Heisenberg XY interaction. The exception is the linear chain but even this case shows a much smaller fidelity on average compared to the Ising model counterpart.

\subsection{\label{sec:subsection3} Anisotropic Coupling}
\begin{figure*}[t!]
    \centering
    \includegraphics[width=\linewidth]{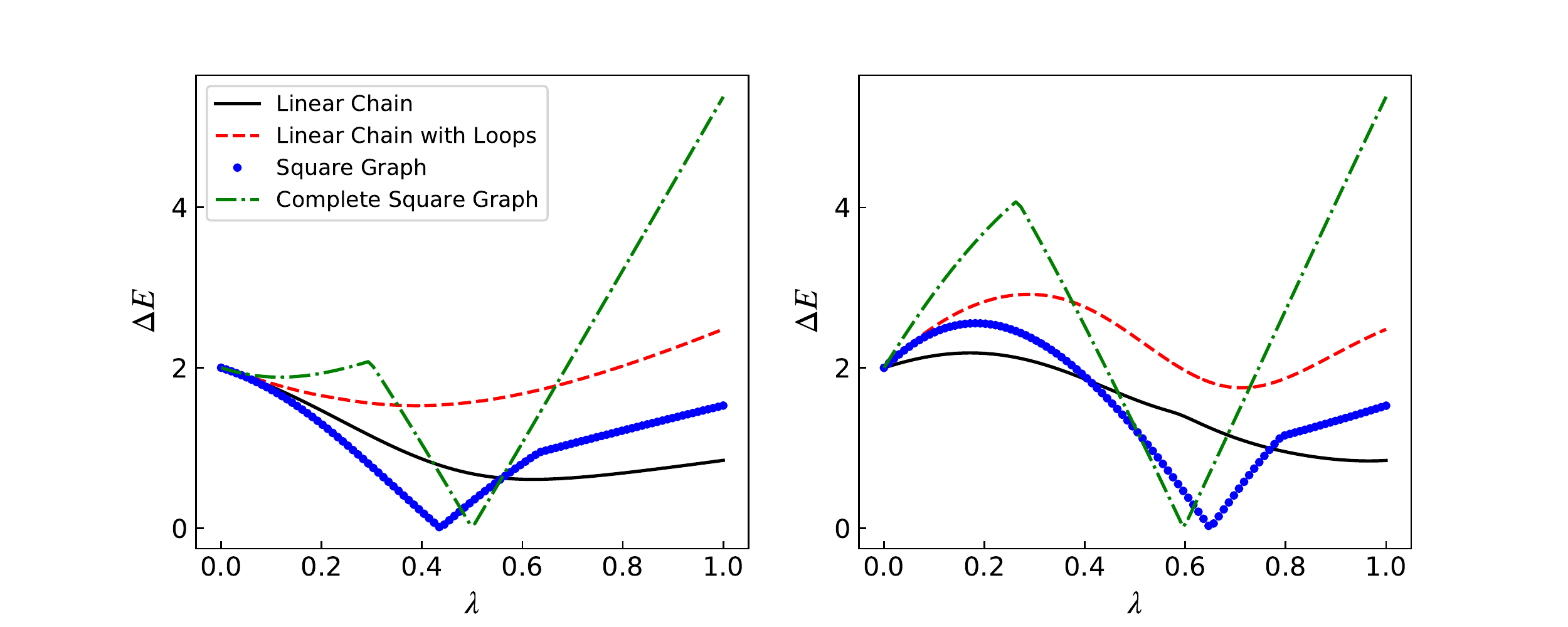}
    \caption{(Color online) Energy gap~($\Delta E$) in the anisotropic coupling scheme. $\lambda$ is the adiabatic control parameter with black solid lines representing the linear chain, red dashed lines for the linear chain with loops, blue dots for the square graph, dash dotted green lines for the complete square graph. The panel on the left shows the results following the model of Eq.~(\ref{eq:adia1}). On the other hand panel on the right follows the model of Eq.~(\ref{eq:trigger}). $J_{ij}$ coefficients are $-1$ and $h_i$ field terms are $1$, trigger Hamiltonian strength $g$ was taken 2. }
    \label{fig:anisoEn}
\end{figure*}

The final model that we give the results of is the Dzyaloshinskii-Moriya interaction model. We check every figure of merit again to see the performance of AQC in this context. Our observation is that the energy gap calculations that are in Fig.~\ref{fig:anisoEn} speak most about the performance of the anisotropic model.

Before going over the energy gaps behavior we should explain why the other parameters like coherence and fidelity failed to give a better picture. The square graph and complete square graph both show discontinuities regarding their entropy, coherence, magnetization, and fidelity. Even though they show very high fidelity after those jumps, it isn't necessarily a positive aspect in terms of adiabatic computation. The reason for that can be best explained with their energy gap. Here on the left panel of Fig.~\ref{fig:anisoEn}, both square graph and complete square graph experience a zero energy gap during the process. This creates a problem since we are working with finite time and those discontinuities indicate a requirement for infinitesimal steps to achieve adiabaticity (see Refs.~\cite{LandauZenner,BoundsFor}) as it is very easy to jump into an exited state. 

To overcome this zero energy gap difficulty we sought alternative models and one such way recommended was the addition of an extra term to Hamiltonian. Those are referred to as trigger terms or steering terms~\cite{Steering,Trigger}. The trigger Hamiltonian's purpose is to control the energy gaps of the spin network for favor of the applicability of the quantum adiabatic transformation. The specific choise of the trigger Hamiltonian is not unique and would change for different models. For the anisotropic exchange model we employed a trigger Hamiltonian of the form
\begin{equation}
    H_t = g\sum_{i\neq j}^NJ_{ij}S^x_iS^x_j,
    \label{eq:trig1}
\end{equation}
where, $g$ is the control parameter for trigger Hamiltonian's strength. With this extra term the adiabatic process becomes
\begin{equation}
    H(\lambda) = \lambda H_1 + (1-\lambda)H_0 + \lambda(1-\lambda)H_t.
    \label{eq:trigger}
\end{equation}
Here, the trigger term vanishes at the beginning and the end of the process, ensuring we will have the same ground state as Eq.~(\ref{eq:adia1}). The results of this addition can be seen in the right panel of Fig~\ref{fig:anisoEn}. There we can see the 0 energy gap problem continues for the square and complete square graph even with the presence of the added term. Also, increasing or decreasing $g$ didn't help either to improve results. However, the extra term helps to increase the energy gap in general, positively affecting chain graphs in terms of their performance. We should also note that the original works we refer to for this method mainly use the Ising model so there was no guarantee to fix the anisotropic case. 

We can also talk about an alternative trigger term for the anisotropic model. Without justifying its use we tried
\begin{equation}
    H_t = g\sum_{i\neq j}^NJ_{ij}S^y_iS^y_j,
\end{equation}
in place of Eq.~(\ref{eq:trig1}). The results with the same $g=2$ and $J_{ij}=-1$ parameters show that the energy gap of the complete square graph is lifted. Then, we can say that together with the vanishing energy gap problem and the model-specific Hamiltonian terms that can lift energy gaps can be of theoretical interest in future works.

We conclude that square graphs are inefficient choices for adiabatic computation in the anisotropic model relative to linear chains. Especially the linear chains with loops showing decent fidelity results. The main difference in fidelity between linear chains seems to be caused by the higher entropy of the linear chain with loops.

\subsection{\label{sec:subsection4} Inhomogeneities}

So far, the couplings that we used for different models and different network connections were the same. But those structures might not have homogeneous couplings, since the qubits might have different displacements from each other or other factors that affect coupling strengths. We want to see how much does inhomogeneous couplings affect the performance of any quantum network. Coupling values are defined with a vector $J=[J_{12},J_{13},J_{14},J_{23},J_{24},J_{34}]$ in the example of the complete square graph of Fig.~\ref{fig:graphs}. 

\begin{figure}[h!]
\includegraphics[width=\linewidth]{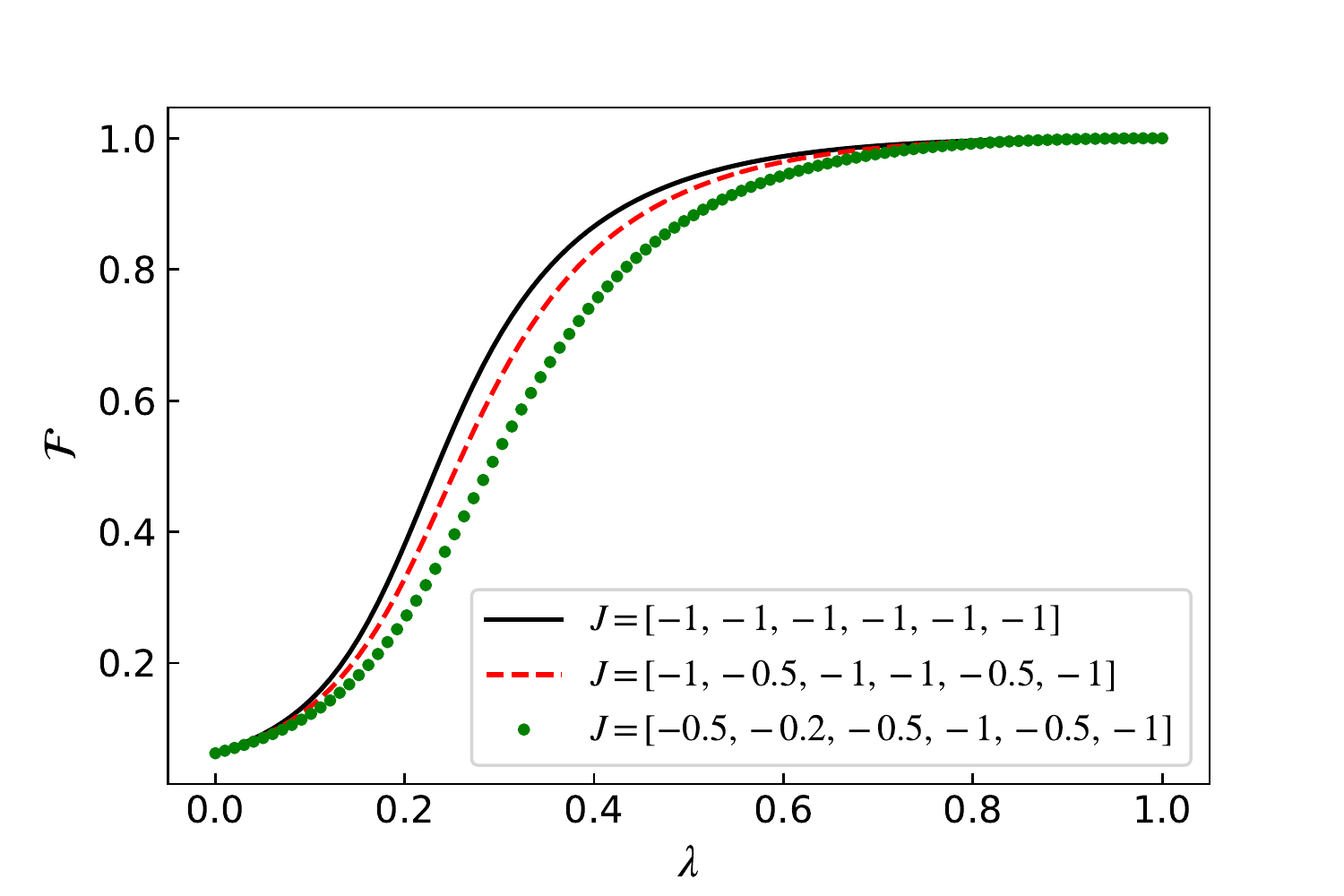}
\caption{\label{fig:inhom}(Color Online) Fidelity~($\mathcal{F}$) results for the complete square graph with the Ising model coupling scheme. $\lambda$ is the adiabatic control parameter and black solid lines represent homogeneous $J_{ij}$ couplings. Red dashed lines and green dots represent the increasing amount of inhomogeneities with $J_{ij}$ arrays $[-1,-0.5,-1,-1,-0.5,-1]$, $[-0.5,-0.2,-0.5,-1,-0.5,-1]$ given respectively. }
\end{figure}

Results are presented in Fig.~\ref{fig:inhom} regarding the complete square graph with Ising model couplings.
We observe that homogeneous couplings have higher fidelity on average. This is true for any graph structure set up with Ising model linear chains and square graphs alike. Inhomogeneities that we show in Fig.~\ref{fig:inhom} have negative effects on the efficiency of Ising model quantum networks.

When we consider Heisenberg model, the complete square graph shows much better fidelity results unlike what we see with Fig.~\ref{fig:hei}. This was also true for the linear chain with loops and the square graph improving fidelity results but inhomogeneities affect the linear chain worse. 

For the case of the anisotropic model, similar to the Heisenberg model, improvements over fidelity are apparent for both square-type graphs but linear chains suffer. We should also note that the energy gap problem that we have mentioned with Fig.~\ref{fig:anisoEn} for square graphs got resolved with inhomogeneous couplings. However, this fix isn't about inhomogeneities but the strength of the couplings as we have seen much better results for anisotropic case for square graphs taking $J_{ij}=-0.5$ instead of $J_{ij}=-1$. As a result we see that for Heisenberg XY model and anistropic model weaker coupling strengths increase the performance for the square type graphs. 

\section{\label{sec:conc}Conclusion}
In summary, we have examined the problem of the effects of different quantum network structures over the efficiency and possibility of successful adiabatic quantum computation. Three models for quantum network structures were considered; the Ising model, Heisenberg XY model, and anisotropic exchange model. We used the method of exact diagonalization to find eigenvalues with computational aid. We used those eigenvalues with their corresponding ground states in the computational basis to measure five figures of merits. Those include energy gap, entanglement entropy, mean magnetization, $l_1$ norm coherence, and fidelity. 

Our major conclusions are that the graph networks that define the level of interaction between qubits heavily affect the possibility of efficient adiabatic quantum computation. But the significance of those effects changes with the choice of model. We have found that the Ising model allows for efficient computation for any network structure but Heisenberg and anisotropic coupling schemes favor the linear spin chains. An apparent reason for the failure of square or ladder-type graph structures was the much higher entanglement entropy that they produce. This fact shows that entanglement entropy while required to run computations, can have unwanted effects depending on the model and graph structure. We have also checked the effect of trigger Hamiltonians to improve the results for square graphs with the anisotropic couplings. It appears that we should select trigger terms according to the model as the $S^yS^y$ coupling trigger term worked much better than the $S^xS^x$ one for the complete square graph. Looking at the inhomogeneous coupling cases convinced us that weaker couplings increase the energy gap minimum from zero for the anisotropic case. This was also true for the Heisenberg model networks, as both square graphs had better fidelity results. Linear chains, on the other hand, suffer from inhomogeneities for all models. Another minor finding was that the adiabaticity was easy to achieve with the systems we have considered. Energy gap requirements were satisfied except the anisotropic model square graphs and speed in which we change $\lambda$ didn't have a dramatic effect.
 
From a fundamental perspective, our results illuminate the interplay of the graph structure of the interacting systems for quantum annealing. Our study signifies the limitations that those structures can impose on adiabatic quantum computers and therefore should be noted for their design as well. We suggest using linear chains if the annealing is to be done with Heisenberg XY or anisotropic models unless the couplings are weak compared to the usual case with the Ising model.


\section*{Acknowledgements}

We would like to thank O\u{g}uzhan G\"{u}rl\"{u} and A. Levent Suba\c{s}ı for fruitful discussions.



\bibliography{ref}

\end{document}